\documentclass[usenatbib]{mn2e}
\usepackage{graphicx}

\title[Results from WASP0 I: Analysis of the Pegasus Field]
{Results from the Wide Angle Search for Planets Prototype (WASP0) I:
Analysis of the Pegasus Field}
\author[S.R. Kane et al.]{Stephen R. Kane$^1$, Andrew Collier Cameron$^1$,
Keith Horne$^1$, David James$^{2,3}$,
\newauthor Tim A. Lister$^1$, Don L. Pollacco$^4$, Rachel A. Street$^4$,
Yiannis Tsapras$^5$\\
$^1$School of Physics \& Astronomy, University of St Andrews, North Haugh,
St Andrews, Fife KY16 9SS, Scotland\\
$^2$Department of Physics \& Astronomy, Vanderbilt University, Nashville,
TN 37235, USA\\
$^3$Laboratoire d'Astrophysique, Observatoire de Grenoble, BP 53, F-38041,
Grenoble, Cedex 9, France\\
$^4$School of Mathematics and Physics, Queen's University, Belfast,
University Road, Belfast, BT7 1NN, Northern Ireland\\
$^5$School of Mathematical Sciences, Queen Mary University of London,
Mile End Road, London, E1 4NS, UK}

\begin{document}

\maketitle

\begin{abstract}

WASP0 is a prototype for what is intended to become a collection of
wide-angle survey instruments whose primary aim is to detect extra-solar
planets transiting across the face of their parent star. The WASP0
instrument is a wide-field (9-degree) 6.3cm aperture F/2.8 Apogee 10 CCD
camera (2Kx2K chip, 16-arcsec pixels) mounted piggy-back on a commercial
telescope. We present results from analysis of a field in Pegasus using the
WASP0 camera, including observations of the known transiting planet around
HD 209458. We also present details on solving problems which restrict the
ability to achieve photon limited precision with a wide-field commercial
CCD. The results presented herein demonstrate that millimag photometry can
be obtained with this instrument and that it is sensitive enough to detect
transit due to extra-solar planets.

\end{abstract}

\begin{keywords}
methods: data analysis -- stars: variables -- stars: individual (HD 209458)
-- planetary systems
\end{keywords}

\section{Introduction}

In the search for extra-solar planets, indirect methods have been used to
detect the more than 100 planets discovered thus far (e.g.,
\citet{may95,mar96,del98}). Of the indirect methods, the use of transits
\citep*{bor85} is rapidly developing into a strong and viable means to
detect extra-solar planets. A transit occurs when the apparent brightness
of a star decreases temporarily due to an orbiting planet passing between
the observer and the stellar disk, leaving a photometric signature of the
planet in the lightcurve of the parent star.

The radial velocity surveys have shown that roughly 7\% of sun-like
(F5--K5) stars in the solar neighbourhood are orbited by a jupiter-mass
companion in the orbital range 0.035--4.0 AU. Moreover, 0.5\%--1\% of
sun-like stars in the solar neighbourhood have a jupiter-mass companion in
a 0.05 AU (3--5 day) orbit \citep{lin03}. If the orbital plane of these
``hot jupiters'' is randomly oriented, approximately 10\% (averaged over
all sun-like stars) will transit the face of their parent star as seen by
an observer. Thus roughly 1 sun-like star in 1000 will produce detectable
transits due to an extra-solar planet. Since this transit method clearly
favours large planets orbiting their parent stars at small orbital radii,
a large sample of stars must be monitored in order to detect statistically
meaningful numbers of transiting planets.

Over the entire sky, roughly 1000 stars brighter than 13th magnitude
should be exhibiting lightcurve dips due to transiting jupiter-mass
extra-solar planets. The photometric accuracy required to reliably detect
a transiting planet in a 3--5 day orbit is considerably higher than that
typically necessary for CCD photometry of variable stars, since starlight
dims by about 1\% for a few hours during each orbital period. Because
thousands of lightcurves must be studied to find one that exhibits
transits, a signal-to-noise (S/N) ratio of at least 6--10 is needed to
reliably detect the transits. Since the cost of detecting transiting
planets can be dominated by the cost of a CCD, it is useful to test if
commercial large-format CCD cameras are able to achieve milli-magnitude
accuracy on 1 hour timescales, and thus sufficient accuracy to be able to
reliably detect transits. Success would show that a large number of similar
cameras could be relatively cheaply deployed in order to monitor the entire
sky for transits.

We present first results from the Wide Angle Search for Planets prototype
(hereafter WASP0), a wide-field instrument used to search for planetary
transits. We describe the data reduction methods and show that the
required accuracy can be achieved with this instrument provided that
sources of systematic errors are treated appropriately. Observations of
the known transiting planet, HD 209458b, are presented as well as a
selection of other stars in the field which exhibit transit-like
lightcurves. The WASP0 instrument was developed as a proof-of-concept
instrument similar to Vulcan \citep{bor01} to search for planetary
transits, and as a precursor to SuperWASP \citep{str03}, a more advanced
instrument that has recently been constructed on La Palma, Canary Islands.

\section{WASP0 Hardware}

WASP0 is an inexpensive prototype for SuperWASP, whose primary aim is to
detect transiting extra-solar planets. The WASP0 instrument is a
wide-field (9-degree) 6.3cm aperture F/2.8 Nikon camera lens, Apogee 10
CCD detector (2K $\times$ 2K chip, 16-arcsec pixels) which was built by
Don Pollacco at Queen's University, Belfast. Calibration frames were used
to measure the gain and readout noise of the chip and were found to be
15.44 e$^-$/ADU and 1.38 ADU respectively. Images from the camera are
digitized with 14-bit precision giving a data range of 0--16383 ADUs.
Figure 1 shows the expected WASP0 sensitivity with the main losses
resulting from the CCD quantum efficiency. The instrument uses a clear
filter which has a slightly higher red transmission than blue.

\begin{figure}
  \includegraphics[angle=270,width=8.2cm]{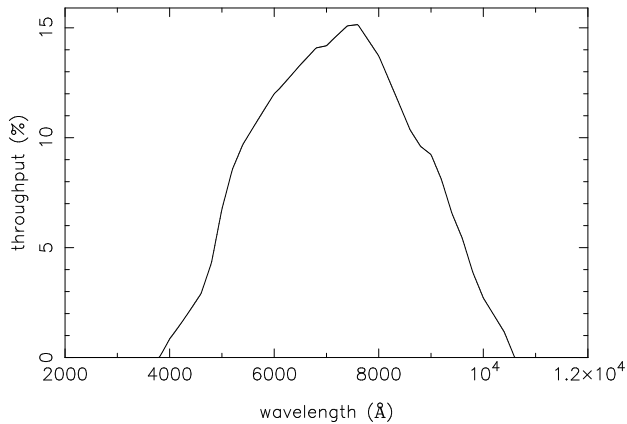}
  \caption{CCD response of WASP0.}
\end{figure}

During its observing run on La Palma, Canary Islands, WASP0 was mounted
piggy-back on a commercial 8-inch Celestron telescope with a German
equatorial mount. More recently at Kryoneri, Greece, WASP0 was mounted on
a 10-inch Meade with a fork equatorial mount.

\section{Observations}

WASP0 has had two successful observing runs at two separate sites. The
first observing run was undertaken on La Palma, Canary Islands during
2000 June 20 -- 2000 August 20. The second observing run took place at
Kryoneri, Greece between 2001 October -- 2002 May. In order to monitor
sufficient numbers of stars for successful planetary transit detection,
a wide field needs to be combined with reasonably crowded star fields.
Observations on La Palma concentrated on a field in Draco which was
regularly monitored for two months. More recent observations from Greece
have largely been of the Hyades open cluster along with further
observations of the Draco field. Observations of the Draco field from La
Palma were interrupted on four occasions when a planetary transit of HD
209458 was predicted. On those nights, a large percentage of time was
devoted to observing the HD 209458 field in Pegasus. This paper details
results from the Pegasus field while the results from the Draco and Hyades
observations will be published elsewhere. The fields centre for each field
monitored by WASP0 is shown in Table 1.

\begin{table}
\caption{Field centres of WASP0 observations (J2000.0)}
\begin{tabular}{@{}lcc}
Field & RA & Dec\\
Pegasus & 22 03 11 & 18 53 04\\
Draco & 17 40 00 & 47 55 00\\
Hyades & 04 24 40 & 17 00 00\\
\end{tabular}
\end{table}

During the night, the WASP0 camera was operated via a series of Visual
Basic scripts which automated the observations and collected both science
and calibration frames. The dead time between exposures was typically the
same as the readout time for the chip (approximately 8 seconds) since the
camera is ready as soon as readout is complete. With the exception of the
Pegasus field observations, science frames generally alternated between
20s and 120s to extend the dynamic range so that brighter stars saturated
in the longer exposure would be unsaturated in the shorter exposure.

\begin{figure}
  \includegraphics[width=8.2cm]{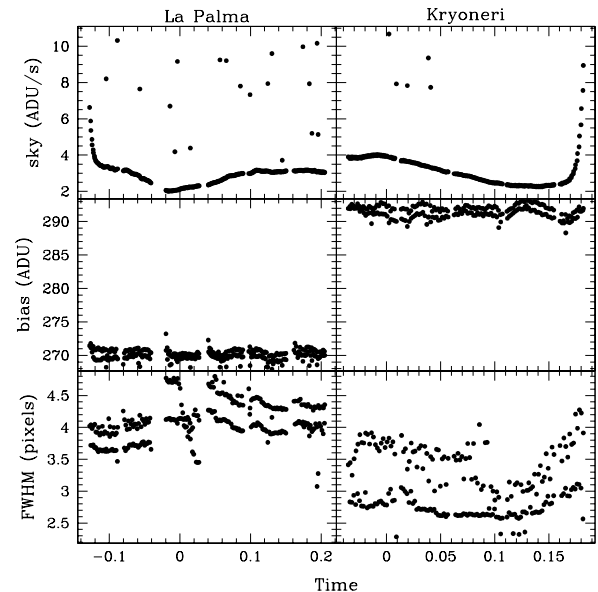}
  \caption{Comparison of observing sites, La Palma and Kryoneri. The time
is expressed as fractional days since local midnight.}
\end{figure}

Figure 2 presents a comparison of the two observing sites used for WASP0
observations. Each of the nights are dark nights and the data are from
observations of the same field. The sky brightness is quite similar
between the two sites, though light pollution from the nearby city of
Corinth combined with the cloud-producing bay make observations from
Kryoneri towards the east difficult. The points of high sky brightness
shown are likely caused by the observer's torch shining too close to the
WASP0 field of view. The middle panels show that the bias level remains
fairly constant and drifts over a range of 10 ADU in the course of each
night. The point-spread function is defined mainly by camera optics rather
than atmospheric seeing due to the large pixel size. Hence the FWHM is
often larger in the 120s than in the 20s exposures due to smearing of the
stellar profiles arising from wind shake and tracking errors. This effect
can be clearly seen in the lower panels of Figure 2.

A problem encountered with observations from La Palma resulted from the use
of the German equatorial mount. A well-known flaw of the German equatorial
mount is that it is impossible to sweep continuously from east to west in
one movement. As the telescope nears the meridian, the telescope must be
swung around to the other side of the mount and the target must be
re-acquired so that observations can continue to the west. The result of
this, besides interruptions to the observing, is that frames in the second
half of the night are rotated 180$^\circ$ relative to frames obtained in
the first half. The fork equatorial mount used at Kryoneri presented no
such problem, allowing continuous observations of targets.

\section{Data Reduction}

During the first two months of WASP0 observations, nearly 150 Gigabytes of
data were obtained. A data pipeline has been developed to reduce this
dataset with a high level of automation. The reduction of wide-field
optical images requires special care since there are many spatially
dependent aspects which are normally assumed to be constant across the
frame. The airmass and even the heliocentric time correction vary
significantly from one side of the frame to the other. The most serious
issues arise from vignetting and barrel distortion produced by the camera
optics which alter the position and shape of stellar profiles. This tends
to be particularly severe in the corners of the image (see Figure 3). Many
of these issues are compounded in the 2000 La Palma data due to the image
rotation caused by the use of the German equatorial mount.

\begin{figure}
  \includegraphics[width=8.2cm]{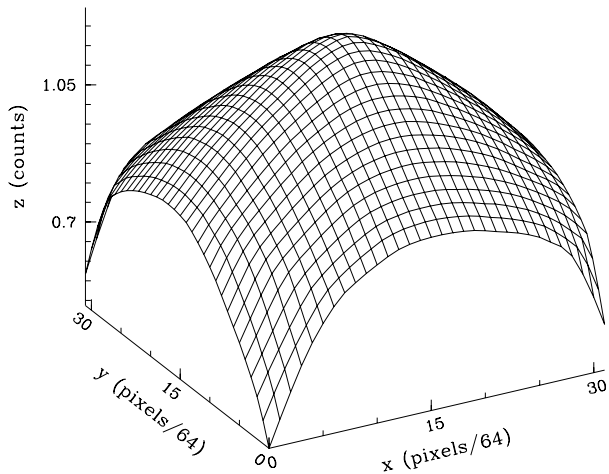}
  \caption{Vignetting pattern across the surface of the frame as measured
from the normalised median flat-field.}
\end{figure}

These problems have been largely solved through the implementation of an
astrometric fit which uses both the Tycho-2 \citep{hog00} and USNO-B
\citep{mon03} catalogues. Rather than fit the variable point-spread
function (PSF) shape of the stellar images, weighted aperture photometry
is used to compute the flux centred on catalogue positions. Post-photometry
calibrations are then applied to remove nightly trends from the data. These
steps will now be described in greater detail.

\subsection{Point-Spread Function Profiles}

The images acquired with WASP0 are normally slightly de-focussed so that
the point-spread functions have a FWHM of a few pixels. Spreading out
the starlight in this way increases the sky background against which
each star must be measured. The hope is that this sacrifice of
statistical accuracy will result in reduced systematic errors that could
arise from undersampling of stellar images by the 15-arcsec pixels.

Due to the wide-field and the optics of the camera, the shape of the PSF
varies with a roughly radial dependence. The photometry package used most
extensively was a modified version of DoPHOT \citep*{sch93}, the most
important modification being the use of the flat-field to calculate a
more adequate noise model for the image. However, DoPHOT still used the
weighted average of the PSF shape parameters when subtracting stars
from the image and so the shape parameters were dominated by the large
number of circular shaped stars in the centre of the frame. Figure 4 (left)
shows an sample of stars from a WASP0 image with the central star being
close to the centre of the image. Figure 4 (right) shows the same stars
after the spatially-independent PSF model has been subtracted. It has also
been discovered that there is a strong colour dependence which is
strengthened by the use of a clear filter during observations.

\begin{figure*}
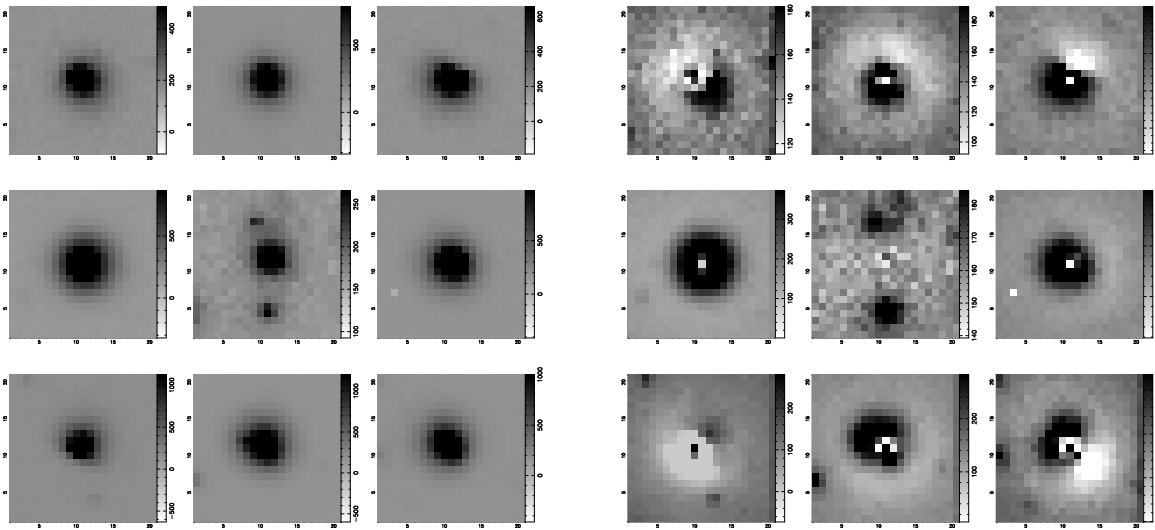

  \begin{center}
    \begin{tabular}{cc}
      \includegraphics[width=7.2cm]{figure04a.ps} &
      \hspace{0.5cm}
      \includegraphics[width=7.2cm]{figure04b.ps} \\
    \end{tabular}
  \end{center}
  \caption{A sample of star images demonstrating the shape of the PSF in
the centre of the image and around the edges. The figure on the right shows
the same stars after a model PSF has been subtracted.}
\end{figure*}

\subsection{Frame Classification and Calibration}

WASP0 FITS headers contain minimal information, so an automated frame
classifier was developed to identify the nature of the frames.
Identification is achieved by statistical measurements and the frames
are classified as one of bias, flat, dark, image, or unknown. Master
calibration frames are then generated after quality control checks are
performed which count the number of saturated pixels.

Since the WASP0 camera has no overscan region, bias frames were taken
at regular intervals during the night. However, it was discovered that a
dead column on the CCD could effectively be used as an overscan region
and so the median value of this strip is subtracted from the frame. Sky
flats were used to flat-field the science frames but were difficult to
obtain due to the 40\% vignetting in the corners relative to the centre
and the short exposures needed to avoid saturation. A shutter time
correction map is applied to the master flat to take into account the
radial and azimuthal structure caused by the shutter blades on
exceptionally short exposures.

\subsection{Astrometry}

The first step in the astrometry procedure is to create an object list
from an image with rough aperture magnitudes using the Starlink {\tt
Extractor} program, which is derived from {\tt SExtractor} \citep{ber96}.
The coordinates of the nominal field centre are used to extract a subset
of the Tycho-2 catalogue, covering an area slightly wider than the
camera's field of view. Star coordinates in the tangent plane are computed
from the Tycho-2 catalogue by gnomonic projection, after transformation of
the catalogue coordinates from mean to observed place. The gnomonic
projection uses initial estimates of the optical axis location on the sky
and the barrel distortion coefficient.

The WASP0 astrometry software automatically cross-identifies a few
dozen bright stars present on both lists, and computes an initial
4-coefficient astrometric solution which describes the translation,
scaling, and rotation of the CCD coordinates in the tangent plane. This
solution allows the cross-identification of many more stars found by
{\tt Extractor} with their Tycho-2 counterparts. The locations of the
optical axis on the sky and on the CCD are refined iteratively,
together with the barrel-distortion coefficient, using the downhill
simplex algorithm. At each iteration a new 4-coefficient astrometric
fit is computed. Once the solution has converged, a final 6-coefficient
fit is computed, which corrects for any small amount of image shear
that may be present.

\begin{figure}
  \includegraphics[width=8.2cm]{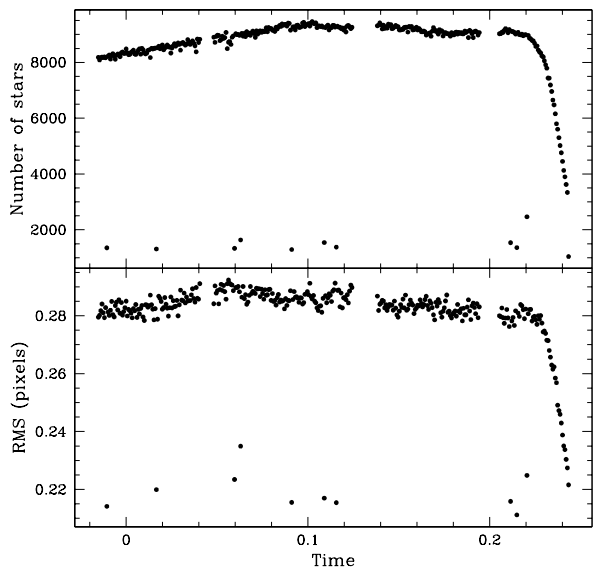}
  \caption{Number of stars used for the astrometric fit and the RMS
of the fit. The time is expressed as fractional days since local
midnight.}
\end{figure}

Figure 5 shows the number of stars used in computing the final astrometric
fit and the associated rms error. On frames of high sky background,
especially noticable during morning twilight, faint stars are excluded from
the fit resulting in a slightly improved rms. Once the solution has been
computed, the much deeper USNO-B catalogue is read and the object
positions within the cameras's field of view are transformed to CCD $x,y$
positions and are written to a catalogue along with the RA and Dec
coordinates, the heliocentric time correction, and the airmass. This
output catalogue is now ready for the photometry stage.

\subsection{Photometry}

Using the USNO-B catalog stars, we excluded circular regions with radii
depending on star brightness to define a skymask. The curvature of the sky
in the wide-field frames presented a new problem in determining local sky
values and hence producing adequate photometry. This was solved through a
sky fitting routine which fits a quadratic surface to the image excluding
circles of tunable radius around the known positions of stars from the
catalogue. In this sense, the deeper USNO-B catalogue resulted in a great
improvement of the fit since faint stars which would artificially raise the
sky level are also excluded when fitting the surface. An iterative process
then ensues during which the sky fit is refined by applying a sigma-clip to
remove cosmic rays and other outliers. Once the final image of the
quadratic surface is produced, the pixels removed from the fit are used to
create a sky rejection mask.

The aperture photometry routine then computes the flux in a circular
aperture of tunable radius centred on the predicted positions of all
objects in the catalogue. The sky background is computed beyond a larger
exclusion radius after subtracting the quadratic sky surface and excluding
pixels defined in the sky rejection mask. Since the aperture is centred on
the actual star position, the weights assigned to pixels lying partially
outside the aperture are computed using a Fermi-Dirac-like function. This
is tuned to drop smoothly from 1.0 at half a pixel inside the aperture
boundary to 0.0 at half a pixel outside the boundary. The weights of
these edge pixels are renormalised to ensure that the effective area of
the aperture is $\pi r^2$ where $r$ is the aperture radius in pixels.
This acts to ``soften'' the edges of the aperture to take into account
the large pixel size. The resulting photometry still contains time and
position dependent trends which we removed by post-photometry
calibration.

\subsection{Post-Photometry Calibration}

The post-photometry calibration of the data is achieved through the use
of a post-photometry calibration code. This constructs a theoretical model
which is then subtracted from the data leaving residual lightcurves. The
residuals are then fitted via an iterative process to find systematic
correlations in the data.

As well as fitting for time dependence, airmass, and colour; the
theoretical model can optionally include a polynomial fit of degree
$\le 3$ to take spatial variance into account. A theoretical model for
the predicted instrumental magnitude of star $s$ consisting of time
$\Delta m_{t(t)}$, colour $\Delta m_{c(t)}$, airmass $\Delta m_{a}$, and
quadratic spatial dependence takes the form
\begin{eqnarray}
  m(s,t) & = & m_s(s) \nonumber \\
  & + & \Delta m_t(t) + \Delta m_c(t) \ \eta_c(t)
  + \Delta m_a \ \eta_a(s,t) \nonumber \\
  & + & \Delta m_x(t) \ \eta_x(s,t)
  + \Delta m_y(t) \ \eta_y(s,t) \nonumber \\
  & + & \Delta m_{xx}(t) \ \eta_{xx}(s,t)
  + \Delta m_{yy}(t) \ \eta_{yy}(s,t) \nonumber \\
  & + & \Delta m_{xy}(t) \ \eta_{xy}(s,t)
\end{eqnarray}
where $m_s(s)$ is the instrumental magnitude of star $s$, the $\Delta m$
terms represent magnitude corrections for various systematic errors, and
the $\eta$ terms are the corresponding dimensionless basis functions. For
example, $\Delta m_t(t)$ accounts for a time-dependent sensitivity. The
extinction coefficient $\Delta m_a$ scales the basis function
\begin{displaymath}
  \eta_a(s,t) = a(s,t) - a_0.
\end{displaymath}
to correct the airmass $a(s,t)$ for star $s$ at time $t$ to a fiducial
airmass $a_0 = 1$. Similarly, the colour term $\Delta m_c(t)$ corrects to
a fiducial colour index, and the spatial terms correct to the centre of the
chip. Iterations of the chosen calibration model are considered to have
converged if the magnitude difference in applying the model is less than an
arbitrarily small value (usually $10^{-6}$). RMS vs magnitude plots are
available to evaluate the improvement by applying the model and the
de-trended lightcurves can then be further analysed for variable and
transit signatures.

\subsection{Transit Detection Algorithm}

The final stage in the WASP0 data processing is the search for planetary
transit signatures in the stellar lightcurves. There have been a variety
of methods (eg., \citet*{def01,doy00,kov02}) discussed on the topic of
automating transit searches. Two of the important issues for such methods
are the reduction of computational time to a reasonable value and the
optimisation of the model to avoid false positive detections. The method
used here is a matched-filter algorithm which generates model transit
lightcurves for a selected range of transit parameters and then fits them
to the stellar lightcurves.

The transit model used for fitting the lightcurves is a truncated cosine
approximation with four parameters: period, duration, depth, and the time
of transit midpoint. The search first performs a period sweep with a fixed
duration over all the lightcurves. The advantage of fixing the duration to
a reasonable value and then scanning for multiple transits is that it
dramatically reduces the number of false positive detections by avoiding
single dip events. The search is refined by performing a duration sweep on
those stars which are fitted significantly better by the transit model
compared with a constant lightcurve model. A transit S/N statistic is 
calculated for each lightcurve based on the resulting reduced $\chi^2$ and
$\Delta \chi^2$. The folded lightcurves of these stars are examined
individually to assess the fit of the transit model. The transit detection
algorithm is discussed in more detail in Scetion 5.3.

\section{Results}

Presented in this section are the results from four nights of monitoring
the Pegasus field. These nights were planned to coincide with predicted
transits by the known extra-solar planet orbiting HD 209458
\citep{cha00,hen00}.

\subsection{Photometric Accuracy}

The upper panel of Figure 6 shows the RMS vs magnitude diagram achieved
by the aperture photometry previously described after correction of
systematic errors. The lower panel shows the same RMS accuracy divided
by the predicted accuracy for the CCD. The data shown include around
7800 stars at 311 epochs from a single night of WASP0 observations and
only includes those stars for which a measurement was obtained at 90\%
of epochs. The night in question was 8th August, 2000 during which 50
second exposures were taken and, although the night was clear, was during
bright time. The location of HD 209458 on the diagram is indicated by a
5-pointed star. The black curve indicates the theoretical noise limit
with the 1-$\sigma$ errors being shown by the dashed lines either side.
It should be noted that this theoretical noise limit assumes that optimal
extraction using PSF fitting has been used, with a CCD noise model given
by
\begin{equation}
  \sigma^2 = \left( \frac{\sigma_0}{F(x,y)} \right)^2 +
  \frac{\mu(x,y)}{G \ F(x,y)}
\end{equation}
where $\sigma_0$ is the readout noise (ADU), $G$ is the gain (e$^-$/ADU),
and $F(x,y)$ is the flat-field.

\begin{figure}
  \includegraphics[width=8.2cm]{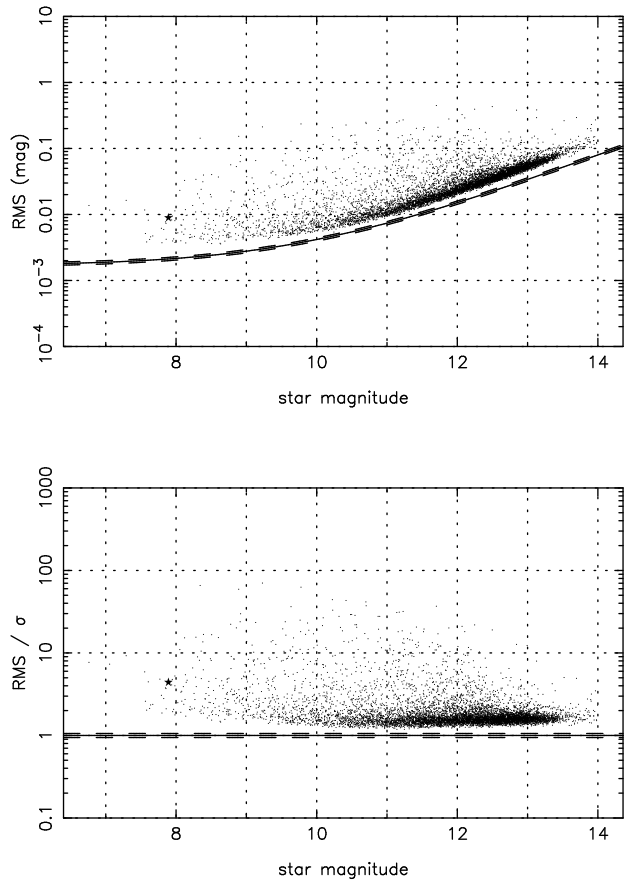}
  \caption{Photometric accuracy vs magnitude diagram from the night of
8th August, 2000 with the location of HD 209458 shown as a 5-pointed star.
The upper panel shows the RMS accuracy in magnitudes in comparison with
the theoretical accuracy predicted based on the CCD noise model. The
lower panel is the ratio of the observed RMS divided by the predicted
accuracy.}
\end{figure}

There is a notable offset between the RMS accuracy and the predicted
accuracy shown in Figure 6 due to sky photon noise and scintillation
noise. It is also worth noting that a number of stars exhibit a magnitude
jump, which occurs when the field is rotated 180$^\circ$, due to the
change in shape of the PSF. This is particularly severe in the case of
faint stars with bright neighbours. Nonetheless, an RMS scatter of 4
millimags is achieved for stars brighter than magnitude 9.5, rising
smoothly to 0.01 mag at magnitude 11.0. This demonstrates that WASP0 is
able to achieve the millimag accuracy needed in order to be sensitive to
planetary transits due to extra-solar planets.

\subsection{HD 209458}

The field in Pegasus containing HD 209458 was monitored closely on four
nights of predicted transits; 25th July, 1st August, 8th August, and
15th August, 2000. Each of these nights were relatively clear with the
exception of 1st August which was frequently interrupted by cloud. The
exposure times on each of these nights was 5, 30, 50, and 50 seconds
respectively. In order to match with the frames from the second night,
the 5 second frames were rebinned in groups of six to create
equivalent 30 second exposures.

Figure 7 shows the HD 209458 lightcurves on each of the four nights in
which the planetary transit is clearly visible. This successfully tests
the capability of the WASP0 system to detect small-scale deviations in
stellar lightcurves.

\begin{figure}
  \includegraphics[width=8.2cm]{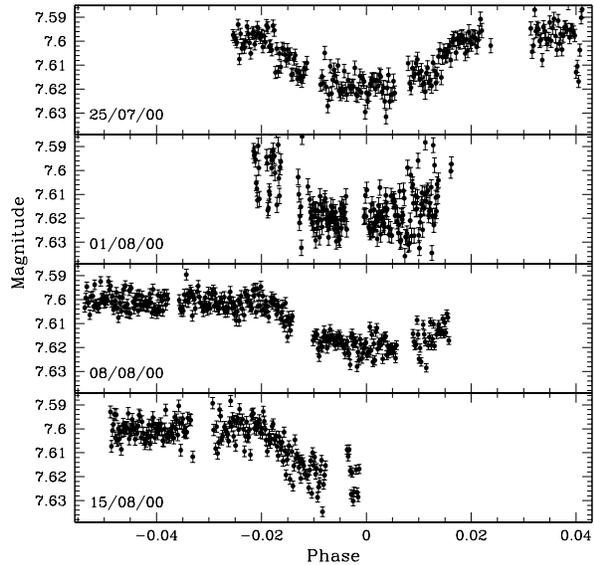}
  \caption{HD 209458 lightcurves from four nights of Pegasus observations
showing the transit of the parent star by the planet HD 209458b.}
\end{figure}

\subsection{Planetary Transit Search}

The field monitored in Pegasus provides an ideal test dataset for the
transit detection algorithm described earlier, since there is known to
be at least one transiting planet in the field. Passing the lightcurves
through the algorithm with a period range of between 3 and 4 days and with
a fixed duration of 3 hours produces the plots shown in Figure 8, with the
location of HD 209458 on each diagram indicated by a 5-pointed star. It
can be seen that HD 209458 is well separated from the majority of stars
in each of these diagrams. These results are of course already biased
towards a HD 209458 planet detection since we only observed the star on
nights on which a transit was predicted to occur.

\begin{figure*}
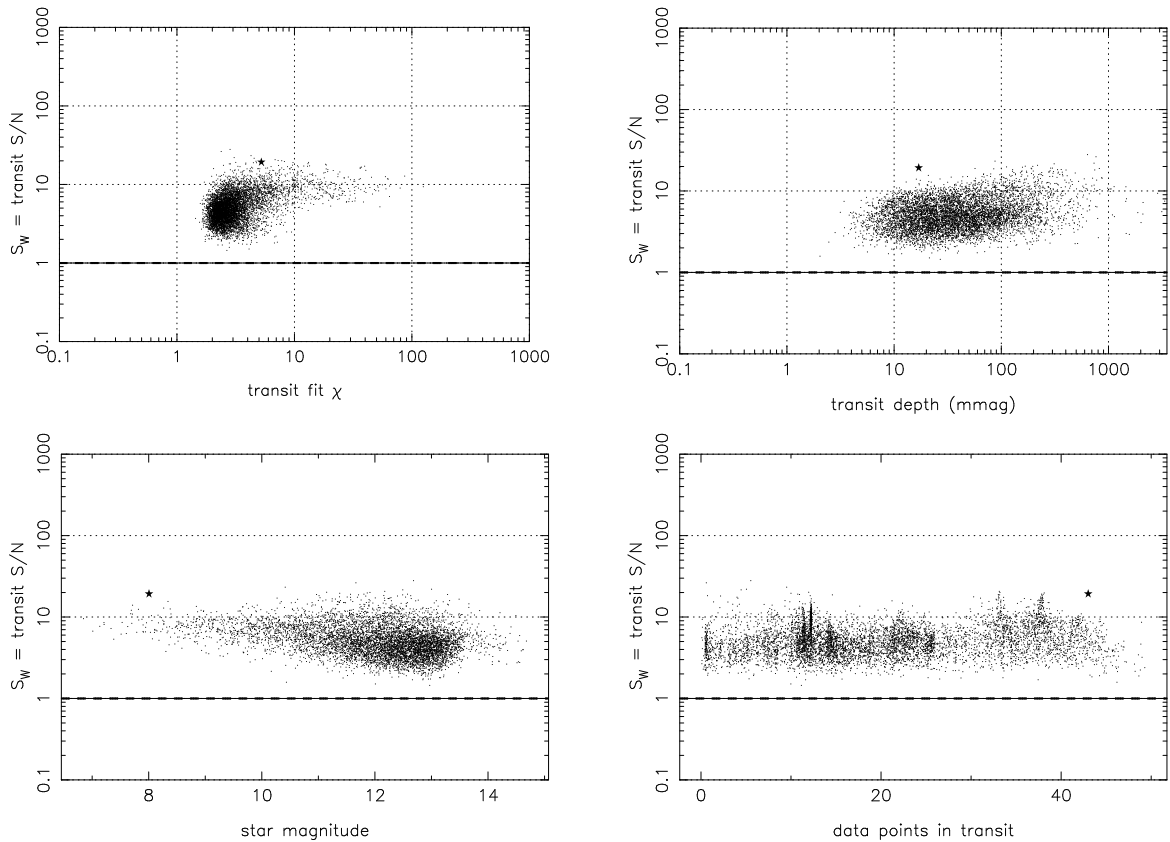

  \begin{center}
    \begin{tabular}{cc}
      \includegraphics[angle=270,width=7.2cm]{figure08a.ps} &
      \hspace{0.5cm}
      \includegraphics[angle=270,width=7.2cm]{figure08b.ps} \\
      \includegraphics[angle=270,width=7.2cm]{figure08c.ps} &
      \hspace{0.5cm}
      \includegraphics[angle=270,width=7.2cm]{figure08d.ps} \\
    \end{tabular}
  \end{center}
  \caption{Results of a passing the data through the transit detection
algorithm with a period range of between 3 and 4 days with a fixed duration
of 3 hours. Each of the plots are plotted against the transit S/N which
measures the ``goodness-of-fit'' of each staller lightcurve to a transit
model.}
\end{figure*}

Each of the four plots shows various star characteristics plotted against
the transit S/N, $S_W$, which is defined as
\begin{equation}
  S_W^2 = \frac{\Delta \chi^2}{\chi_{\mathrm{min}}^2 / (N - f)}
\end{equation}
where $N$ is the number of data points, $f$ is the number of free
parameters, and $\Delta \chi^2$ is given by
\begin{displaymath}
  \Delta \chi^2 = \chi_{\mathrm{constant}}^2 - \chi_{\mathrm{transit}}^2.
\end{displaymath}
The errors in the individual lightcurves are rescaled by a factor that
forces the $\chi_{\mathrm{min}}^2 / (N - f)$ for each lightcurve to be
equal to unity. The $S_W$ statistic shown in equation 3 is used
consistently throughout the transit detection algorithm, including the
first pass in which the transit duration is fixed. As can be seen from
the plots in Figure 8, this transit S/N is an effective means for sifting
transit candidates from the data.

\begin{figure*}
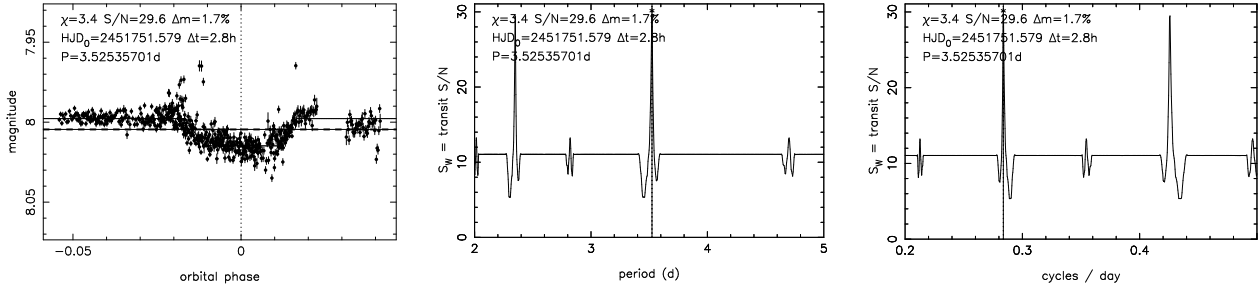

  \begin{center}
    \begin{tabular}{ccc}
      \includegraphics[angle=270,width=5.2cm]{figure09a.ps} &
      \hspace{0.0cm}
      \includegraphics[angle=270,width=5.2cm]{figure09b.ps} &
      \hspace{0.0cm}
      \includegraphics[angle=270,width=5.2cm]{figure09c.ps} \\
    \end{tabular}
  \end{center}
  \caption{Results of a refined transit model fit to the data obtained of
HD 209458. The left panel shows the folded data with the best fit model.
The middle panel shows the periodogram from a period sweep ranging from 2
to 5 days. The right panel is a frequency plot which shows the number of
cycles/day.}
\end{figure*}

An arbitrary $S_W$ (transit S/N) threshold may be used to isolate stars
with possible transits for further analysis. In Figure 8, for example,
stars which yield $S_W > 10$ would be an ideal group to investigate
further. For stars with multiple transits, the folded lightcurves may be
examined which also gives a quick estimate of the transit depth. Using
colour information from the Tycho-2 and USNO-B catalogues, these transit
depths can be translated into Jupiter radii and hence it can be determined
whether the star may classified as a transit candidate and suitable for
further follow-up.

The transit model was refined for HD 209458 by allowing the transit
duration vary between 2 and 4 hours and expanding the period range to
between 2 and 5 days with much higher time resolution. These results can
be seen in Figure 9. The fit parameters for the planetary transit of
HD 209458 are a depth of $\Delta m = 1.7\%$, a duration of $\Delta t =
2.790 \pm 0.015 \ \mathrm{hours}$, and a period of $P = 3.5239 \pm
0.0003 \ \mathrm{days}$. The fitted duration in our model is slightly
unreliable since data covering the egress of the transit is missing on
most nights (see Figure 7).

The range of periods over which a transit search can be conducted depends
upon the sampling and duration of the survey. In particular, large gaps in
the sampling can lead to cycle-count ambiguity since $n$ unobserved
transits may have occurred during the gaps. This is clear from the
periodogram shown in Figure 9 which shows aliases that differ from the real
period by 1 cycle per 7 days. Figure 9 also contains a frequency plot which
shows that the alias periods are equally spaced by 1 cycle per 7 days.

\subsection{Transit-like Events and Variable Stars}

\begin{figure*}
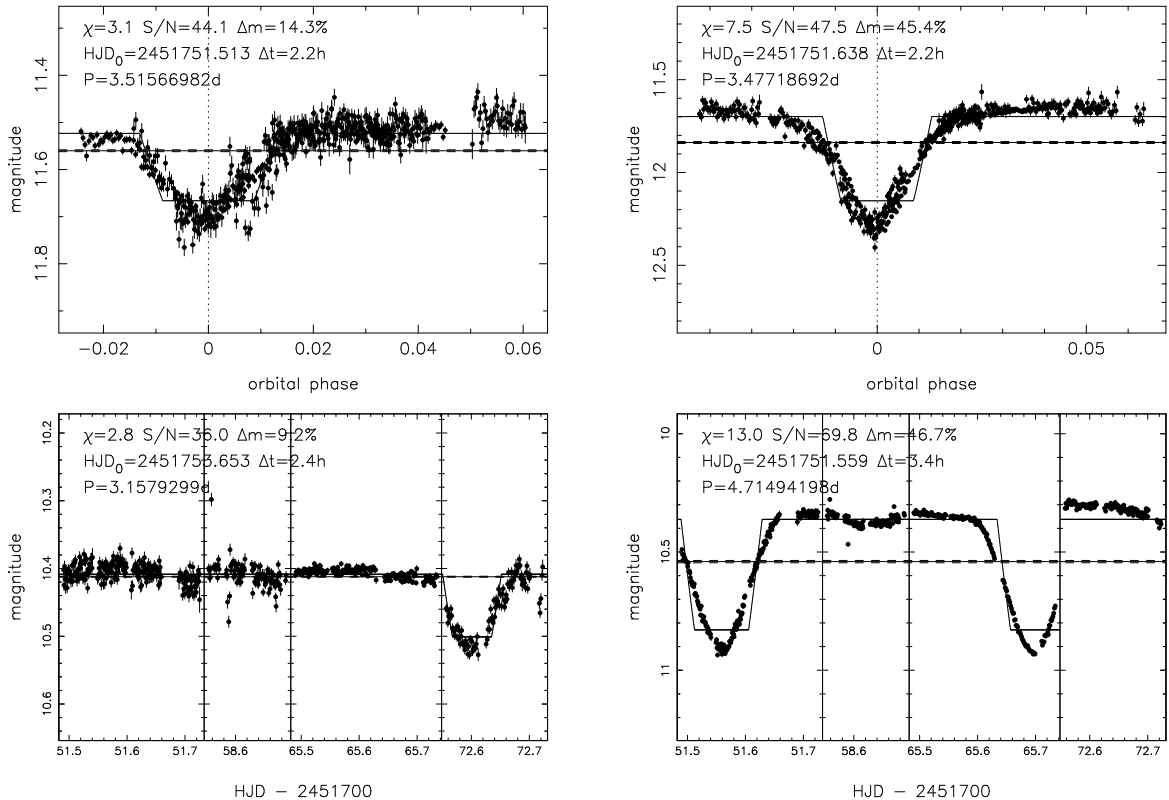

  \begin{center}
    \begin{tabular}{cc}
      \includegraphics[angle=270,width=7.2cm]{figure10a.ps} &
      \hspace{0.5cm}
      \includegraphics[angle=270,width=7.2cm]{figure10b.ps} \\
      \includegraphics[angle=270,width=7.2cm]{figure10c.ps} &
      \hspace{0.5cm}
      \includegraphics[angle=270,width=7.2cm]{figure10d.ps} \\
    \end{tabular}
  \end{center}
  \caption{Four example stars exhibiting transit-like or variable
behaviour. The top two panels are show folded lightcurves and the
bottom two panels are unfolded.}
\end{figure*}

The monitoring program of WASP0 is ideal for detecting many kinds of
stellar variability. As a result, the Pegasus field data have yielded
lightcurves of stars whose variable nature was previously unknown. Shown
in Figure 10 are four examples of stars in the field which exhibit
variable behaviour identified by the transit detection algorithm. Clearly
each of these stars are either too deep or too variable to be a genuine
planetary transits and are likely to to be eclipsing or grazing eclipsing
binaries. A comprehensive study of new variable stars discovered in the
Pegasus field will be published elsewhere.

\section{Conclusions}

We have described an inexpensive prototype survey camera designed to
monitor a wide-field on-sky area and hunt for transiting extra-solar
planets. Large datasets have so far been obtained of fields in Draco and
Hyades with some additional nights monitoring Pegasus at predicted transit
times of the known extra-solar planet HD 209458b. A pipeline has been
developed which makes use of the Tycho-2 and USNO-B catalogue to provide
an astrometric solution for each frame. The transformed source list is
then used to perform weighted aperture photometry at the location of each
object in the catalogue.

The resulting analysis of WASP0 Pegasus data presented here demonstrates
that this instrument is able to achieve millimag photometry and hence the
necessary precision required to detect transit events due to extra-solar
planets. A transit detection algorithm has been described and has shown
that it is effective at extracting stars with exhibit transit-like
behaviour from the bulk of the dataset.

It is unfortunate that the WASP0 camera only observed using a clear
filter. This means that multi-colour follow-up observations of transit
candidates, needed to distinguish other possible physical scenarios,
will be required to be sought after elsewhere.

This prototype has successfully served as a proof-of-concept for
SuperWASP, a robotically operated, multi-camera instrument recently
constructed on La Palma. The primary science goal of the SuperWASP
project is to use the well-sampled lightcurves to detect planetary
transits, but the data will also be used to detect Near-Earth Objects
and optical transients.

\section*{Acknowledgements}

The authors would like to thank Aleksander Schwarzenberg-Czerny for useful
discussions regarding the transit detection algorithm. The authors would
also like to thank PPARC for supporting this research and the Nichol Trust
for funding the WASP0 hardware.


\begin{thebibliography}{}
\bibitem[\protect\citeauthoryear{Bertin \& Arnouts}{1996}]{ber96} Bertin,
E., Arnouts, S., 1996, A\&AS, 117, 393
\bibitem[\protect\citeauthoryear{Borucki, Scargle, \& Hudson}{Borucki et
al.}{1985}]{bor85} Borucki, W.J., Scargle, J.D., Hudson, H.S., 1985, ApJ,
291, 852
\bibitem[\protect\citeauthoryear{Borucki et al.}{2001}]{bor01} Borucki,
W.J., Caldwell, D., Koch, D.G., Webster, L.D., Jenkins, J.M., Ninkov, Z.,
Showen, R., 2001, PASP, 113, 439
\bibitem[\protect\citeauthoryear{Charbonneau et al.}{2000}]{cha00}
Charbonneau, D., Brown, T.M., Latham, D.W., Mayor, M., 2000, ApJ, 529, L45
\bibitem[\protect\citeauthoryear{Delfosse et al.}{1998}]{del98} Delfosse,
X., Forveille, T., Mayor, M., Perrier, C., Naef, D., Queloz, D., 1998,
A\&A, 338, L67
\bibitem[\protect\citeauthoryear{DeFa\"y, Deleuil, \& Barge}{DeFa\"y et
al.}{2001}]{def01} DeFa\"y, C., Deleuil, M., Barge, P., 2001, A\&A, 365,
330
\bibitem[\protect\citeauthoryear{Doyle et al.}{2000}]{doy00} Doyle, L.R.,
et al., 2000, ApJ, 535, 338
\bibitem[\protect\citeauthoryear{Henry et al.}{2000}]{hen00} Henry, G.W.,
Marcy, G.W., Butler, R.P., Vogt, S.S., 2000, ApJ, 529, L41
\bibitem[\protect\citeauthoryear{H{\o}g et al.}{2000}]{hog00} H{\o}g, E.,
et al., 2000, A\&A, 355, L27
\bibitem[\protect\citeauthoryear{Kov\'acs, Zucker, \& Mazeh}{Kov\'acs et
al.}{2002}]{kov02} Kov\'acs, G., Zucker, S., Mazeh, T., 2002, A\&A, 391,
369
\bibitem[\protect\citeauthoryear{Lineweaver \& Grether}{2003}]{lin03}
Lineweaver, C.H., Grether, D., 2003, ApJ, 598, 1350
\bibitem[\protect\citeauthoryear{Marcy \& Butler}{1996}]{mar96} Marcy, G.,
Butler, R., 1996, ApJ, 464, L147
\bibitem[\protect\citeauthoryear{Mayor \& Queloz}{1995}]{may95} Mayor, M.,
Queloz, D., 1995, Nat, 378, 355
\bibitem[\protect\citeauthoryear{Monet et al.}{2003}]{mon03} Monet, D.G.,
et al., 2003, ApJ, 125, 984
\bibitem[\protect\citeauthoryear{Schechter, Mateo, \& Saha}{Schechter et
al.}{1993}]{sch93} Schechter, P.L., Mateo, M., Saha, A., 1993, PASP, 105,
1342
\bibitem[\protect\citeauthoryear{Street et al.}{2003}]{str03} Street, R.A.,
et al., 2003, ASP Conf. Series, Vol. 294, Scientific Frontiers in Research
on Extrasolar Planets, eds. D. Deming \& S. Seager, p. 405
\end{thebibliography}
\end{document}